\documentstyle[eqsecnum,aps,twocolumn]{revtex}

\begin{document}
\author{A.K.Avetissian, K.Z.Hatsagortsian, G.F.Mkrtchian, Kh.V.Sedrakian}
\address{Department of Theoretical Physics, Plasma Physics Laboratory, Yerevan State
University, \\
1, A. Manukian, 375049 Yerevan, Armenia\\
E-mail: mkrtchian@sun.ysu.am\\
Fax: (3741) 151-087}
\title{NON LINEAR COMPTON SCATTERING OF STRONG LASER RADIATION ON CHANNELED
PARTICLES IN A CRYSTAL}
\maketitle

\begin{abstract}
A version for intense $\gamma $-ray radiation based on the multiphoton
scattering of strong laser radiation on relativistic particle beam channeled
in a crystal  is proposed. The scheme is considered when the incident laser
beam and charged paricles beam are counter-propagating and the laser
radiation is resonant to the energy levels of transversal motion of
channeled particles.
\end{abstract}

\section{INTRODUCTION}

As is known channeling occurs if a charged particle enters a crystal at an
angle to a crystallographic axis or plane smaller than Linchard angle $%
\theta _L=\sqrt{2U_0/\varepsilon }$, where $U_0$ is the depth of a
transverse potential well, and $\varepsilon $ is the particle's energy \cite
{kumb}. The spontaneous radiation of channeled particles \cite{wal}, \cite
{kuma} has some significant properties that open possibilities for
implementation of short-wave radiation sources since due to their large
Doppler shift and the high oscillation frequency in the channel ($\Omega
\sim 10^{14}-10^{16}s^{-1}$for particles energies $\varepsilon \sim
10GeV-10MeV$) particles emit quanta mainly in the X-ray and $\gamma $-ray
domain with an intensity much higher than the intensity of other types of
radiation (see \cite{BAZb} and references therein).

The spontaneous radiation of channeled particles has been comprehensively
studied both theoretically and experimentally (see, e.g., Refs \cite{kumb},%
\cite{BAZb}), while induced channeling radiation, that is radiation in the
presence of external electromagnetic wave (EMW), has been studied mainly in
the linear regime of interaction \cite{5}-\cite{12}. To achieve a
considerable amplification in the single pass non-linear regime of X-ray
amplification has been studied in \cite{av1}, but these investigations show
that implementation of short-wave coherent radiation sources due to
stimulated channeling radiation is far from being realized yet. One reason
is that because of short lifetime of a particle transverse-motion levels the
length of coherent interaction of a channeled particle with EMW is quite
short (e.g., of the order of one micrometer at positron energies $\sim 10MeV$%
) compared to the interaction length in other versions of Free Electron
Laser (such as the undulator and Cherenkov lasers). There is an another
problem connected with the controlling of the channeled particles
overpopulation\cite{13}.

Two component laser assisted schemes for radiation enhancement have also
been studied. One of those is based on stimulated photon scattering by
channeled particles \cite{11}. Since absorption of a photon by a channeled
particle is a resonant process, the cross section of photon scattering by a
channeled particle is $10^4$ times larger than the free electron Compton
scattering cross section \cite{10}. Nevertheless, the gain of a EMW in
stimulated Compton scattering on channeled particles does not exceed the
gain of stimulated emission in the channel from the initially inverse
populated states.{\rm \ }The second scheme concerns the quantum mode of
interaction, that is the coherent radiation of quantum modulated beam at the
frequency of the stimulating wave\cite{av2} and its harmonics \cite{av3}. 

In this paper we investigate multiphoton scattering of strong laser
radiation on relativistic particle beam channeled in a crystal which can
serve as a possible scheme for $\gamma $-ray generation. The scheme is
considered when averaged potential for a plane channeled particles is good
enough described by the harmonic potential. Then it is assumed that the
incident laser beam and charged particles beam are counter-propagating and
the laser radiation is resonant to the energy levels of transversal motion
of channeled particles. In the result of the multiphoton scattering the hard
quanta are generated. The discussed scheme has several advantages in respect
to the known ones. First of all, the cross section of the process is
resonantly enhanced in respect to the Compton scattering process. At the
second, the multiphoton processes arise at the much lower laser intensities
than in the case of the Compton scattering. Besides, the scheme enables to
use the forward channeling radiation arising due to transitions from short
living, high excited states of a particle to the ground state, that could
not be achievable in the process of spontaneous channeling radiation.

This paper is organized as follows. In Section II the wave function of a
plane channeled particle in a EMW is obtained. In Sec.\ III within the scope
of Quantum Electrodynamics the spectral intensity of multiphoton Compton
scattering in a strong laser field is obtained. The  first-order Feynman
diagram, where the electron/positron lines correspond to the wave functions
in the strong laser field is calculated and the resonant case of interaction
is discussed.

\section{Wave Function of a Plane Channeled Particle in the Field of
Transverse Electromagnetic Wave}

We will consider the case when averaged potential of the crystal for a plane
channeled particle is good enough described by the harmonic potential 
\begin{equation}
U(x)=\kappa \frac{x^2}2  \label{1}
\end{equation}
for plane channeled positron 
\begin{equation}
\kappa =\frac{2U_0}{d^2}
\end{equation}

where $U_0$ is the transverse potential holes depth, and $d$-is the
interspace distance \cite{BAZb}. For plane channeled electrons the
approximate potential is actually not harmonic, but for the high energies it
can be approximated by (\ref{1}). As it is known, for the channeled
particles the depth of the potential hole is $U_0<<E$, where $E$ is the
particle energy. The spin interaction which is $\symbol{126}{\bf %
\bigtriangledown }\cdot U(x)$~ is again less than $E$. For this reason the
transversal motion is good enough describes by the Schr\"{o}dinger equation 
\cite{BAZb} with effective mass $m_{ef}=E_{\parallel }$ (the natural units $%
\hbar =c=1$ will be used throughout this paper), where $E_{\parallel }=\sqrt{%
p_z^2+m^2}$ is the energy of longitudinal motion, $m$ is the particle mass.

On the other hand, the spin interaction can play a role in a spontaneous
radiation process if the radiated photons energy is $\omega \symbol{126}E$.
But if the particles energy is not enough high, i.e.

\begin{equation}
E<<\frac{m^2}{E_{\perp }},  \label{2}
\end{equation}
where $E_{\perp }$ is the energy of transversal motion, then $\omega <<E$
and the spin effects are not substantial. Let us mention that (\ref{2}) can
be re-written in the following way

\begin{equation}
E_{\perp }<<\frac E{\gamma ^2},  \eqnum{2.2$\prime $}  \label{eq:mynum}
\end{equation}
(where $\gamma $- is the Lorenz factor) which is the condition allowing to
neglect the impact of the transversal oscillations on the longitude motion.
We will consider that not only but also the change of the transversal
motions energy which is resulted from the interaction with the external EM
field is small

\begin{equation}
\Delta E_{\perp }<<\frac E{\gamma ^2}.  \label{3}
\end{equation}
As there is a relation between the total $E$ and transverse energy changes $%
\Delta E_{\perp }$ :

\[
\Delta E=\gamma ^2\Delta E_{\perp } 
\]
then due to the Doppler-shift of the emitted radiation, the condition (\ref
{3}) actually restricts the total energy change

\begin{equation}
\Delta E<<E_0,  \eqnum{2.3$\prime $}  \label{eq:myn}
\end{equation}
which means that anyway for the frequency of the external electromagnetic
radiation $\omega _0<<E$. For this reason in the considering process of
interaction of the channeled particles with the external EM radiation the
spin interaction will not play any role. So we will use the Klein-Gordon
equation

\begin{equation}
\left[ i\frac \partial {\partial t}-U\left( x\right) \right] ^2\Psi =\left[
\left( \widehat{{\bf p}}-e{\bf A}\right) ^2+m^2\right] \Psi ,  \label{4}
\end{equation}
where $e$ is the particle charge and 
\begin{equation}
{\bf A}=\left\{ A_0\cos \omega _0(t+nz),0,0\right\}  \label{7}
\end{equation}
is the vector potential of the plane EMW. Here $n$ is the crystal refracting
index on $\omega _0$, So, as the external EM field depends only on the $\tau
=t+nz$ then raising from the problem symmetry, the wave function can be
found in the following form:

\begin{equation}
\Psi ({\bf r},{\bf t})=f(x,\tau )\exp \left[ ip_yy+ip_zz-iEt\right]
\label{8}
\end{equation}

Taking into account (\ref{3}) we can consider $f(x,\tau )$ as a slowly
varying function of $\tau $ and neglect the second derivative compared with
the first order. So for $f(x,\tau )$ we will have the following equation:

\[
\left[ \partial _{xx}+2E_{\parallel }(E_{\perp }-U(x))+2i\widetilde{p}%
\partial _\tau \right. 
\]
\begin{equation}
\left. -2iA(\tau )\partial _x-e^2A^2(\tau )\right] f=0  \label{9}
\end{equation}
where

\[
\widetilde{p}=E+np_z. 
\]
In Eq. (\ref{9}) transverse and longitudinal motions are not separated. But
after a certain Unitary transformation in the equation for the transformed
function the variables are separated \cite{av3} and for wave function we
obtain

\[
\Psi =\frac 1{\sqrt{2\Pi }}\exp \left\{ i\Pi _yy+i\Pi _zz-i\Pi t\right\} 
\]
\[
\times \exp \left\{ -i\frac{\widetilde{\omega }(\widetilde{\omega }^2+\Omega
^2)}{8E_{\parallel }\Delta ^2}e^2A_0^2\sin 2\omega _0\tau -ix\frac{\Omega ^2}%
\Delta eA_0\cos \omega _0\tau \right\} 
\]
\begin{equation}
\times V_s\left[ x+\frac{\widetilde{\omega }}{E_{\parallel }\Delta }eA_0\sin
\omega _0\tau \right] \text{,}  \label{10}
\end{equation}
where the state of the particle in EM fields (\ref{1}, \ref{7}) is
characterized by the average energy and momentum ( ''quazimomentum'')
defining via free particle energy-momentum by the following equations.

\[
\Pi _y=p_y,\quad \Pi _z=p_z-n\frac{\widetilde{\omega }^2}{4\widetilde{p}%
\Delta }e^2A_0^2, 
\]

\begin{equation}
\Pi =E+\frac{\widetilde{\omega }^2}{4\widetilde{p}\Delta }e^2A_0^2.
\label{11}
\end{equation}
Here 
\begin{equation}
\widetilde{\omega }=\frac{\widetilde{p}}{E_{\Vert }}\omega _0;\qquad \Delta =%
\widetilde{\omega }^2-\Omega ^2  \label{11b}
\end{equation}
and 
\[
V_s(x)=\frac 1{\pi ^{\frac 14}}\sqrt{\frac \chi {2^ss!}}\exp \left[ -\frac{%
\chi ^2x^2}2\right] H_s(\chi x),\quad 
\]
\[
\chi =\sqrt{E_{\parallel }\Omega },\quad \Omega =\sqrt{\frac \kappa {%
E_{\parallel }}} 
\]
are the wave functions of the harmonic oscillator with Hermit polynomials $%
H_s(\chi x)$. In \ref{10} it is assumed that initial state of the channeled
particle is$\left\{ p_y,p_z,s\right\} $ (before the interaction with EMW).

\section{Compton Scattering on Channeled Particles in a Crystal}

\subsection{Feynman first order diagram}

As we saw in Sec. II $\Pi _y$, $\Pi _z$ and $s$ are the quantum numbers
(neglecting spin interaction) describing the state of a particle moving in
EM fields (\ref{1}, \ref{7}). It is clear that between them there will be a
spontaneous transitions which causes spontaneous radiation. The spontaneous
radiation may be considered by the theory of perturbation. In this case
first order Feynman diagram describes spontaneous radiation where wave
functions (\ref{10}) correspond to electron/positron lines. The probability
amplitude of transition from the state $\left\{ \Pi _{0y},\Pi
_{0z},s_0\right\} $ to the state $\left\{ \Pi _y,\Pi _z,s\right\} $ with
emission of a photon with the frequency $\omega $ and momentum ${\bf k}$
will be \cite{Akh}. 
\begin{equation}
M_{s_0s}^{(\nu )}=e\sqrt{4\pi }j_\mu \frac{e_{0\nu }^{\mu *}}{\sqrt{2\omega }%
}  \label{12}
\end{equation}
where $e_{0\upsilon }^{\mu *}$ is the four-dimensional polarization vector,
index $\nu $ corresponds to the emitted photons of two possible
polarizations ($\nu =1,2$). Here 
\begin{equation}
j^\mu =\int d^4xj_{f,i}^\mu ({\bf r},t)e^{i(\omega t-{\bf kr})}  \label{13}
\end{equation}
where $j_{f,i}^\mu ({\bf r},t)-$is the four-dimensional transition current.

As is known the polarization vector may always be chosen in a way that $%
e_0^\mu =(0,{\bf e}_0);{\bf e}_0{\bf k}=0$ (three dimension gage). The
differential probability calculated in unit volume and unit time will be 
\begin{equation}
dW_{s_0s}^{(\nu )}=\frac{\left| M_{s_0s}^{(\nu )}\right| ^2}{L_yL_zT}\frac{d%
{\bf k}d\Pi _yd\Pi _z}{(2\pi )^3}  \label{14}
\end{equation}
where $L_y,L_z$ are quantization length $T$ is the interaction time. If we
are not interested in the dependence of process on the photons polarization
then the probabilities must be summed by all possible polarizations 
\begin{equation}
\left| M_{s_0s}\right| ^2=\left| M_{s_0s}^{(1)}\right| ^2+\left|
M_{s_0s}^{(2)}\right| ^2  \label{15}
\end{equation}
Taking into consideration that transition current $j^\mu $ satisfies to the
continuity equation 
\[
{\bf jk}=\omega j_0 
\]
the probabilities (\ref{15}) can be presented in the following form

\begin{equation}
\left| M_{s_0s}\right| ^2=\left| {\bf j}\right| ^2-\left| j_0\right| ^2
\label{16}
\end{equation}
that corresponds to the summation by the photons polarizations for $\nu
=1,...4$ in general case.

As long as in our case $\omega <<E$ and the condition (\ref{3}) must be
satisfied then the transition currents may by calculated using the solution
of the Klein-Gordon equation. The transition currents will be

\begin{equation}
{\bf j}_{f,i}=i\left( \Psi _i{\bf \nabla }\Psi _f^{*}-\Psi _f^{*}{\bf \nabla 
}\Psi _i\right) -2e{\bf A}\Psi _i\Psi _f^{*}  \label{17}
\end{equation}

Using this expression for transition currents and taking into account (\ref
{13}), (\ref{14}) and (\ref{16}) we will arrive to the following expression
for the differential probability

\[
dW_{s_0s}=\frac{e^2}{8\pi \omega \Pi \Pi ^{^{\prime }}}\sum_{\ell =-\infty
}^\infty W_{s_0s}^{(\ell )}\delta (\Pi _z-\Pi _{0z}+k_z+\ell n\omega ) 
\]
\begin{equation}
\times \delta (\Pi _y-\Pi _{0y}+k_y)\delta (\Pi -\Pi _0+\omega -\ell \omega
)d{\bf k}d\Pi _yd\Pi _z  \label{18}
\end{equation}

In general the expressions of the partial probabilities are very complicated
and we will not bring here. The $\delta $ functions presenting in the
expression (\ref{18}) for differential probability express the quazimomentum
and quazienergy conservation laws in the given process. Different $\ell $
correspond to different partial processes with fixed photon numbers and $%
W_{s_0s}^{(\ell )}$ are the partial probabilities. Let us find the emitted
photon's energy rising from the conservation laws. Taking into account (\ref
{3}) and $\omega <<E$ we will have the following expression for $\omega $

\begin{equation}
\omega =\frac{1+n\widetilde{v}_{0z}}{1-\frac{{\bf k}\widetilde{{\bf v}}%
_{0\parallel }}\omega }\left[ \ell \omega _0+\Omega ^{\prime }(s_0-s)\right]
\label{20}
\end{equation}
where 
\[
\Omega ^{\prime }=\frac \Omega {1+n\widetilde{v}_{0z}} 
\]
and

\[
\widetilde{{\bf v}}_{0\parallel }=\frac{{\bf \Pi }_{\Vert }}{\Pi _0}
\]
is the mean longitudinal velocity. When $\Omega =0$ and $n=1$ the formula (%
\ref{20}) is reduced to the well known Compton effect one for the scattered
frequency (neglected quantum recoil).

It follows from (\ref{20}) that $\ell >0$ corresponds to the multiphoton
absorption and $\ell <0$ to the multiphoton emission of a wave quanta. It is
noteworthy to mention that in the nonlinear Compton process on free
electrons \cite{Nik} only the case of $\ell >0$ , i.e. the multiphoton
absorption process in the strong EM wave takes place. In contrary, at the
scattering of a strong wave on channeling particles the multiphoton emission
($\ell <0$) of quanta of EMW takes place.

\subsection{Resonance case}

Let us consider the resonance case which is of more interest and expression
for differential probability may be simplified. We will consider the case
when

\begin{equation}
\frac{\left| \widetilde{\omega }_0-\Omega \right| }\Omega <<1;\quad \frac{%
\left| \widetilde{\omega }-\Omega \right| }\Omega <<1  \label{21}
\end{equation}
where $\widetilde{\omega }_0$ and $\widetilde{\omega }$ are initial and
final Doppler shifted frequencies (\ref{21}). Besides we will assume that

\begin{equation}
\xi \equiv \frac{eA_0}m>>\left| \delta \right| ;\quad \delta =\frac{%
\widetilde{\omega }_0-\Omega }\Omega  \label{22}
\end{equation}
Here $\xi $ is the relativistic invariant parameter of the wave intensity.

Considering (\ref{21}) it is possible that despite $\xi <<1$ but the
condition (\ref{22}) may be satisfied.

To obtain the total cross section of nonlinear scattering the expression (%
\ref{18}) must be summed by all discrete states of transverse motion in the
channel. After integrating by $\Pi _y$ and $\Pi _z$, then summing by $\ell $%
, using the $\delta $ functions, and taking into account the (\ref{21}), (%
\ref{22}) for differential cross section we will have

\[
dW=\frac{m^2e^2}{2\pi \omega \omega _0\Pi \Pi ^{^{\prime }}}\left[ -\Lambda
_0^2(N)\right. 
\]

\begin{equation}
\left. +\left( \frac \xi {2\delta }\right) ^2\left[ \Lambda _1^2(N)-\Lambda
_0(N)\Lambda _2(N)\right] \right] d{\bf k}  \label{23}
\end{equation}
Here 
\begin{equation}
\Lambda _r(N,\alpha ,\beta )=(2\pi )^{-1}\int_{-\pi }^\pi d\theta \cos
^r\theta \exp \left[ i\left( \alpha \sin \theta -\beta \sin 2\theta -N\theta
\right) \right]  \label{24}
\end{equation}
are known functions \cite{Nik} and represent non linear processes in the
field of linear polarized wave (multiphoton Compton effect, pair production,
etc.).

\begin{equation}
\alpha =\xi \frac{2mk_x\widetilde{\omega }_0}{E_{\Vert }\Delta \Delta _0}(%
\widetilde{\omega }_0\widetilde{\omega }-\Omega ^2)  \label{25}
\end{equation}
\begin{equation}
\beta =\xi ^2\frac{m^2(\widetilde{\omega }_0-\widetilde{\omega })}{8E_{\Vert
}\Delta \Delta _0}(\widetilde{\omega }_0\widetilde{\omega }-\Omega ^2)
\label{26}
\end{equation}
and $N$ is fixed by the conservation law which in the resonant case is 
\begin{equation}
\omega =\frac{1+n\widetilde{v}_{0z}}{1-\frac{{\bf k}\widetilde{{\bf v}}%
_{0\parallel }}\omega }N\omega _0,\quad  \label{28}
\end{equation}

The formula (\ref{23}) defines the spectral intensity of an one-photon
emission (if product to $\omega $) in the crystal at simultaneously
nonlinear ''Compton'' scattering of a strong EM wave on the channeled
particle at the resonance. Instead of parameter nonlinearity $\xi ^2$ in the
Compton effect on free electrons the effective nonlinearity in the
channeling process is determined by the resonance parameter $\left( \frac \xi
{2\delta }\right) ^2$, increasing the cross sections of the multiphoton
''Compton'' scattering. For the actual cases $\delta \sim 10^{-2}\div 10^{-1}
$ \cite{BAZu}, and consequently the parameter of nonlinearity increases 
\symbol{126}$10^3$ times. As the number of absorbed photons should be
restricted by condition $N\omega _0<<U_0$ ( $U_0$ being the depth of the
channeling potential well) to avoid the dechanneling effects, so for forward
radiation of a particle with the energy $E_{\Vert }\sim 50MeV$, maximum of
emitted quanta energies up to $\hbar \omega \sim 1MeV$ are achievable.

\acknowledgements
We would like to thank Prof. H.K. Avetissian for valuable discussions during
the work under the present paper. This work is supported by International
Science and Technology Center (ISTC) Project No. A-353.

\end{document}